\documentclass[a4paper]{article}
\usepackage{fullpage,amsfonts,url,amssymb,amsmath}
\usepackage{theorem}
\newtheorem{thesis}{Thesis}
{\theorembodyfont{\rmfamily}
\newtheorem{definition}{Definition}[section]}

\begin{document}
\frenchspacing
\title{Some Thoughts on Hypercomputation}
\author{Apostolos Syropoulos\\
        Greek Molecular Computing Group\\
        366, 28th October Str.\\
        GR-671\ 00\ \ Xanthi, GREECE\\
        \texttt{asyropoulos@yahoo.com}}
\date{April 2008}
\maketitle
\begin{abstract}
Hypercomputation is a relatively new branch of computer science that emerged
from the idea that the Church--Turing Thesis, which is supposed to 
describe what is computable and what is noncomputable, cannot possible be true. Because of
its apparent validity, the Church--Turing Thesis has been used to investigate the possible
limits of intelligence of any imaginable life form, and, consequently, the limits
of information processing, since living beings are, among others, information processors. 
However, in the light of hypercomputation, which seems to be feasibly in our universe, one 
cannot impose arbitrary limits to what intelligence can achieve unless there are specific
physical laws that prohibit the realization of something. In addition, hypercomputation 
allows us to ponder about aspects of communication between intelligent beings that have not 
been considered before. 
\end{abstract}
\section{Introduction}
Computer science can be defined as a discipline that studies problems and proposes solutions
to them. A problem is in principle either computable (i.e., mechanically solvable) or 
noncomputable. Those that are in principle computable are either efficiently computable or not. 
This classification is based on the apparent validity of the Church--Turing
Thesis (CTT), a thesis that describes what is computable and what is noncomputable. 
Since not all problems are equally difficult, the disciplines of recursion and complexity theory  
have been developed to deal with the degrees of unsolvability of noncomputable problems and 
the resources required during computation to solve a given problem, respectively. 

Hypercomputation is a new discipline that challenges the validity 
of the CTT and, therefore, seeks to find a new way to characterize problems as computable or 
noncomputable. Obviously, since there is no consensus on what problems can be characterized as 
computable or noncomputable, it makes no sense to discuss about efficiently computable problems,
at least beyond the CTT barrier. The essence of hypercomputation is that we do not know yet 
how to characterize problems as computable and noncomputable. Consequently, what today is 
characterized as noncomputable may easily become computable tomorrow.

Trishank Karthik~\cite{karthik05} has discussed  why, in his opinion, {\em algorithmic} 
communication (i.e., communication whose possibilities is bounded by the validity of the CTT) 
with extraterrestrial intelligence is justified. Karthik very briefly considers hypercomputation 
as a way to seek a solution to the  problem of communication with extraterrestrial intelligence,
nevertheless, he is really reluctant to accept it as a basis for the solution of this problem. In
particular, he wonders if hypercomputation is feasible, then what does it mean to be able to solve 
the {\em halting paradox}, which is a noncomputable problem provided that the CTT is valid. In 
order to answer this question and to provide convincing responses to similar queries, it is 
important to see whether hypercomputation will affect in any way the way we understand our cosmos. 
In particular, we need to see whether the invalidity of the CTT will force us to actually rewrite 
our physics and biology textbooks. In addition, we need to examine how the invalidity of the CTT 
will affect information flow, which lies at the heart of communication. Depending on the outcome 
of these investigations, one may be able to have a better understanding of life as a natural 
phenomenon and to be able to draw conclusions about its future. Naturally, in the end one will 
have a better understanding of the terms {\em computable} and {\em noncomputable}.  

\paragraph{Structure of the paper} We start with a brief of overview of Turing machines, the 
CTT and the halting problem (i.e., Karthik's ``halting paradox'') that is based on~\cite{rogers87}. 
This is followed by a discussion of why the CTT cannot be possible true. Next, we discuss serious
proposals for hypercomputational systems (i.e., feasible systems able to solve {\em noncomputable} 
problems), which are shown not to contradict the laws of physics as we know them today, thus, 
answering the challenge posed by Karthik. Then, there is a discussion about the intellectual 
limits of intelligent beings and the problem of communication with extraterrestrial intelligence 
in the lights of hypercomputation. The paper concludes with a brief summary and some thoughts for 
future research.

\section{On the Church--Turing Thesis}
The CTT says something about the equivalence of algorithms and {\em recursive}
functions (the standard definition will be given below). To put it in a different way, 
it says something about the equivalence of recursive functions and {\em Turing} computable
functions (also the standard definition will be given below). In spite of some efforts to
rigorously define the notion of an algorithm (e.g., see~\cite{horowitz84} for such a definition),
an algorithm is an {\em informal} (i.e., not mathematically exact) notion that, nevertheless, 
plays a central role in computer science. Roughly speaking, an algorithm can be identified 
with an {\em effective} mechanical (i.e., deterministic, book-keeping) procedure applicable 
to a number of symbolic input data that when applied to some symbolic input creates some 
symbolic output. If a number or a function can be computed by an algorithm, then it is 
said to be effectively computable.
 
A Turing machine (TM) is a simple conceptual computing device, which was devised by Alan Mathison 
Turing in the 1930s. In a nutshell, a TM consists of an {\em infinite} tape,\footnote{For 
purists the tape is not infinite but arbitrary extendable to the left and/or to the right.}
which is divided into readable/printable cells, that is, on each cell it is possible to print 
a symbol, or to erase what has been printed on it, or to read the symbol that is printed on 
it. Usually, the symbols are drawn from a set called an {\em alphabet}. In the most simple 
case the alphabet is the set $\{0,1\}$; when $0$ is printed on a cell, this cell is supposed 
to be empty. A scanning head  moves along the tape. When above a cell, the scanning head 
first reads the symbol printed on it, then either erases this symbols, or prints a new symbol,
 or simply leaves the cell intact. The action to be taken depends on the current 
``situation''---what is printed on the current cell, what the scanning head did before and 
what actions are specified in the {\em controlling device}, which is a lookup table that 
specifies what has to be done for each particular case. Initially, the machine's operator 
prints on the machine's tape the input data, which are the data the machine will operate 
on, and sets up the controlling device. It is rather important to say that no assumptions 
are made on how the scanning head recognizes symbols, or how it writes symbols on a cell, 
or how it erases a symbol from a cell. Now, what made the Turing machine the archetypal 
computing device is that it can easily compute a large number of functions and/or numbers, 
which are called {\em Turing} computable. The simplicity of the TM and its computational 
power has prompted many thinkers and researchers to propose that all things in nature are
computational at bottom. Thus, according to this view our brain is a TM and, consequently, 
our mental capabilities are delimited by the capabilities of the TM. This view is known in 
the literature as the {\em computational metaphor}. But if our brains are computational at 
bottom, then one may wonder why the whole universe is not computational at bottom. Indeed, 
the Physics Nobel prize Laureate John Archibald Wheeler was deeply convinced that all things
physical are information-theoretic in origin. This hypothesis is known in the literature 
as the {\em it from bit} hypothesis (see~\cite{wheeler90} for more details). According to 
this hypothesis the most basic ingredients of this universe (whatever this means) are bits, 
that is, chunks of information that are processed by matter. Provided we accept these 
controversial hypotheses, any boundary that delimits the capabilities of TMs, implies 
limits to what can be achieved in the cosmos, in general, and to what we as intelligent 
being can achieve, in particular.

The same time Turing was working on his machine, Stephen Cole Kleene was working on recursive 
functions and managed to define computation as a finite sequence of {\em recursive equations}. 
A (partial) function from the set of natural numbers into natural numbers is {\em recursive}
if it can be represented by an expression formed from certain {\em base functions} and the
operations of {\em composition}, {\em primitive recursion}, and {\em minimization}. The base
functions include the {\em successor} function ($S(x)=x+1$), the {\em zero} function ($z(x)=0$),
and the {\em projection} functions ($U_{i}^{n}(x_{1},\ldots,x_{n})=x_{i}$, where $1\le i\le n$).
Primitive recursion is used to define a function $h(z,x_1,\ldots, x_n)$ from recursive functions
$f(x_1,\ldots,x_n)$ and $g(z,y,x_1,\ldots,x_n)$ by the following equations
\begin{align*}
h(0,x_1,\ldots,x_n) &= f(x_1,\ldots,x_n)\\
h\Bigl(S(z),x_1,\ldots,x_n\Bigr) &= g\Bigl(z,h(z,x_1,\ldots,x_n),x_1,\ldots,x_n\Bigr).
\end{align*}
The operation of composition defines a function $h(x_1,\ldots,x_n)$ from a 
function $f(x_1,\ldots,x_m)$ and $m$ functions $g_i(x_1,\ldots,x_n)$ 
as follows:
\begin{displaymath}
h(x_1,\ldots,x_n)=f\Bigl(g_1(x_1,\ldots,x_n),\ldots,g_m(x_1,\ldots,x_n)\Bigr).
\end{displaymath}
The operation of minimization defines a function $f(x_1,\ldots,x_n)$ from a total recursive function
$g(y,x_1,\ldots,x_n)$ as the ``smallest $y$ such that $g(y,x_1,\ldots,x_n)=0$,'' and is written
\begin{displaymath}
f(x_1,\ldots,x_n)=(\mu y)\Bigl[g(y,x_1,\ldots,x_n)=0\Bigr].
\end{displaymath}
A function (numbers are considered to be constant functions) is computable by a finite sequence of 
recursive equations by substituting numerical expressions for variable symbols through-out an 
equation or ``by the use of one equation to substitute `equals for equals' at any occurreence 
in a second equation'' or by the evaluation of an instance of the successor function. If a 
function is computable in this sense, then it is called {\em Kleene} computable. Interestingly
enough, if a function is Turing computable it is also Kleene computable and vise versa. 
Furthermore, a number of different {\em constructive} formulations of computability turned 
out to be equivalent with each other (e.g., Church's $\lambda$-calculus can be used to compute
exactly the recursive functions). This remark prompted Alonzo Church to formulate a thesis, 
which is now known as the CTT.

Clearly, the CTT is a statement that is supposed to describe what can be computed in general. Although, a number of different formulations of the notion of computability turned out to be equivalent, no one has managed to prove the CTT. Thus, the CTT should be better called a 
{\em hypothesis} or even a {\em conjecture}. A common formulation of the CTT follows:
\begin{thesis}
Every effectively computable function is Turing computable. Alternatively, the effectively 
computable function should be identified with the recursive functions.
\end{thesis}
In other words, a function is computable if it can be computed by a Turing machine, or
alternatively if some function cannot be computed by a Turing machine, then there is no 
way to compute it! However, if one accepts that the CTT is not valid, then (some) 
Turing-noncomputable functions can be computed by some other, more powerful machines. 
In turn, some of these machines
might not be able to compute some other problems, which will be computed by some other even 
more powerful machines, etc. The study of this hierarchy of these machines is the subject 
of the new discipline of hypercomputation.

A {\em universal} TM (UTM) is a special TM that takes as input the 
controlling device of an ordinary TM encoded as an integer and the input of the 
second TM encoded also as an integer. UTMs can simulate the operation of any ordinary TM and 
this is really interesting. Assume that $n$ is an integer encoding some TM $M$ and that 
$m$ is an integer encoding some (meaningful) input for $M$, then it can be proved that 
the {\em halting} function 
\begin{displaymath}
h(n,m)=\left\{\begin{array}{ll}
              1, & \mbox{if $M$ with the given input eventually stops,}\\
              0, & \mbox{otherwise,}
              \end{array}\right. 
\end{displaymath}   
cannot be computed by a UTM. The halting function is the formal expression of the halting 
problem. The later asks whether it is possible to determine if a particular TM
with fixed input will eventually stop. In different words: given some computer program $P$
with some (fixed) input $I$, is there any way to determine whether $P$ will halt?
The halting function is the typical example of an 
noncomputable function. Although it has been proved that this function cannot be computed 
by a TM, the corresponding theorem does not state that it is impossible to 
compute it by some other means. Of course, if one assumes the validity of the CTT, then, 
obviously, the halting function as well as the halting problem are noncomputable. But why 
should we assume that the CTT is actually true?

L{\'a}szl{\'o} Kalm{\'a}r and R{\'o}zsa P{\'e}ter were probably the first thinkers who 
contested the CTT. Kalm{\'a}r~\cite{kalmar59} has argued  that the class of general recursive 
functions is a proper subclass of the class of effectively computable functions. In different 
words, he argued that there are computable functions that are not recursive. 
P{\'e}ter~\cite{peter59} has claimed  
quite the opposite, that is, that the class of recursive functions is
broader that the class of computable functions. Jean Porte~\cite{porte58}  has presented 
another interesting argument against the validity of the CTT. Porte has proved that there 
are general recursive functions $f(z)$ such that, for any general recursive function $g(x)$,
there exist infinitely many numbers $x$ in the range of $f$ such that, for any argument
$z_{0}$ with $f(z_{0})=x$, the number of steps necessary to compute $f(z_{0})$ exceeds
the value of $g(x)$. Assume that $g$ grows very fast, say
\begin{displaymath}
g(x)=100^{100^{100^{x}}}, 
\end{displaymath}
then according to Porte the computation of $f(z_{0})$ is humanly noncomputable since it cannot
be carried out within the life-span of a human being. And from this Porte has concluded that 
the general recursive function $f$ is humanly noncomputable. Elliot Mendelson~\cite{mendelson63}
has noted that humanly computable is not the same as effectively computable. However, according 
to the established view, a function is effectively computable if its value can be computed 
in a finite number of steps, but, obviously, no one has exactly specified what finite 
exactly means. Nevertheless, Vladimir Sazonov~\cite{sazanov95} has made an attempt to specify 
the notion of finiteness. Sazanov, based on the ``observation'' that the number of
elementary particles in the universe is $2^{1000}$, has proposed that $2^{1000}$
is the largest {\em feasible} number (note that such an idea makes sense in the
framework of constructive mathematics only). In other words, Sazanov has identified this number 
with $+\infty$. Consequently, function $f$ is not only humanly noncomputable, but it is
noncomputable in general. If a recursive function is noncomputable, then the 
CTT cannot be possibly valid. Clearly, this is not a bulletproof argument (e.g., the
value of $g$ is far greater than the greatest feasible number), but shows that the
CTT is too vague. Although, these arguments aimed at falsifying the CTT, they do not 
constitute a {\em practical} proposal capable of tackling noncomputable problems. Quite 
naturally, the question is: are there any feasible (conceptual) computing devices more 
powerful than the TM?

{\em Trial-and-error} machines (TAEMs) are feasible conceptual computing devices capable of
tackling noncomputable problems. These machines were discovered independently by Hillary 
Putnam and Mark Gold (see~\cite{putnam65} and~\cite{gold65}, respectively). Basically, 
a TAEM is a TM that can decide whether some element belongs to a subset of the set of natural 
numbers. The machine continuously prints on its tape a sequence of responses (e.g., the digits 
1 and 0) where the last one is always the correct one. In other words, the machine 
is experimenting in order to find the right answer, which is delivered at the end of its 
operation. A TAEM can solve the halting problem, though it may not be able to solve its own 
halting problem, which, however, is an entirely different issue. Although, TAEMs are purely
mathematical ``creatures,'' they  have found unexpected applications. For example, Peter 
Kugel~\cite{kugel86} has proposed a really elegant model of the mind based on TAEMs.

TAE-machines were introduced by Jaakko Hintikka and Arto Mutanen~\cite{hintikka97}. TAE-machines
are a model of computation that is similar to Putnam and Gold's TAEMs. Roughly, a TAE-machine is
a TM with an extra tape, which is called a {\em bookkeeping} tape. Both tapes are read-write 
``storage devices.'' Without loss of generality, one can assume that what appears on the 
bookkeeping tape are equations of the form $f(a)=b$, where $a,b\in\mathbb{N}$. These equations 
are used by the machine to define the function to be computed. In particular, this function is 
computed by the machine if and only if all (and only) such true equations appear on the
bookkeeping tape when the machine has completed its operation. In addition, TAE-machines can solve 
the halting problem and, thus, are more powerful than any TM. And since these machines are
feasible, this is exactly one more reason this author is convinced that the CTT cannot 
possible be true. Nevertheless, these are not the only reasons.

The statement that the equation $x^{n} + y^{n} = z^{n}$  has no non-zero integer solutions 
for $x$, $y$, and $z$ when $n > 2$ is known in the literature as Fermat's last theorem.
As a decision problem this is a semi-decidable problem (i.e., one cannot construct a TM 
that can definitely determine whether this theorem is either true or false; 
see~\cite{rogers87} for more details). In 1995, Andrew John Wiles proved Fermat's last 
theorem by using {\em tricks} from algebraic geometry. Practically, Wiles ``computed''
a recursively enumerable problem. In 2002, Grisha Perelman proved Poincar\'{e}'s 
Conjecture, which is another semi-decidable problem (see~\cite{perelman02,perelman03} 
for details about the problem and its solution). Despite the profound importance of these 
proofs, it seems that their significance has been ignored by the computing world.

In general, a TM is a feasible conceptual computing device that
operates in a brute-force manner, that is, it examines all possible cases one after
another in  order to deliver its result. In the case of Fermat's last theorem, or
Poincar\'{e}'s Conjecture for that matter, a TM has to examine $\aleph_{0}$ 
different cases to verify the statement. Practically, this implies that these problems
are Turing noncomputable! Nevertheless, Wiles' proof, or Perelman's proof for that 
matter, showed that there are methods that involve intuition and {\em intelligence}, 
which can be employed to {\em compute} noncomputable problems. Indeed, hypercomputation is 
also about such methods and how these can be used to solve noncomputable  problems. 

I have not presented a mathematical proof that CTT is false, but then
again no one has proved the opposite. However, I have presented some simple arguments against 
the plausibility of the thesis, which, cannot be easily contested. In addition, these arguments show
that it is logically possible to expect to compute more than what TMs can compute.

\section{Hypercomputation and Physical Reality}
In 1999, Jack Copeland and Diane Proudfoot~\cite{copeland99} coined the term {\em hypercomputation}
in order to describe all conceptual, feasible and infeasible machines that 
transcend the capabilities of the TM, or recursive functions for that matter.  Nevertheless, 
until the day one would come with a realistic scenario or thought experiment that 
would in principle invalidate the CTT, hypercomputation would be classified as yet another 
mathematical curiosity. Fortunately, there are such proposals and I will briefly present 
them in the rest of this section.

Tien Kieu~\cite{kieu03} has proposed  a method to solve Hilbert's tenth problem (HTP), which
is as difficult as the halting problem. In particular, this problem is a decision problem 
that roughly asks whether it is possible to mechanically solve any Diophantine equation
\begin{displaymath}
D(x_1,x_2,\ldots,x_n)=0,
\end{displaymath}
by a finite number of operations in integers, where $n$ is a natural number and $D$ is any 
integer polynomial.  In 1970, Yuri Matiyasevich proved that HTP cannot be decided by 
a TM.  On the other hand, Kieu has claimed to have found a method that {\em mechanically} 
solves HTP. At first sight, the two results seem contradictory. But, both results
can be true, provided the CTT is not valid. Note that I use the term {\em method} instead of 
{\em algorithm} since the later is clearly associated to Turing computable procedures. Kieu's 
method makes use of the {\em adiabatic} theorem of quantum mechanics,\footnote{Interestingly, 
D-Wave Systems, a private company based in Canada, announced the launch of the world's first
``commercial'' quantum computer in 2007. According to the company's founder and chief technology 
officer, their machine is a 16--qubit processor that employs the adiabatic theorem to deliver 
results. Nevertheless, see~\cite{brumfiel07} for a  criticism regarding this announcement.} 
which says something about the way quantum processes evolve. More specifically, it is known 
that Schr\"{o}dinger's equation describes the time-dependence (and space-dependence) of quantum
mechanical systems. In Dirac's notation the time-dependent Schr\"{o}dinger equation, which gives
the time evolution of the state vector $|\Psi(t)\rangle$, has the following form
\begin{displaymath}
i\hbar\frac{d}{dt}|\Psi(t)\rangle=H(t)|\Psi(t)\rangle,
\end{displaymath}
where $i$ is the imaginary unit, $\hbar$ is the reduced Planck's constant, $t$ is time, and 
$H(t)$ is the Hamiltonian operator, which describes the total energy of the system. If we
assume that $H(t)$ varies slowly, then by using the adiabatic theorem we can see how the 
system will evolve. The really difficult task is to find the Hamiltonian. An idea is to find 
the ground state of a Hamiltonian $H_P$ by means of a third Hamiltonian $H_B$. All three 
Hamiltonians should be related through the following time-dependent process:
\begin{displaymath}
H\left(\frac{t}{T}\right)=\left(1-\frac{t}{T}\right)H_{B}+\frac{t}{T}H_{P}.
\end{displaymath}
According to the adiabatic theorem, if the deformation is slow enough, the initial state 
will evolve into the desired ground state with high probability. Of course, Kieu's method 
has a number of steps that must be followed (see, for example,~\cite{kieu03} 
or~\cite{syropoulos08} for more details), but the use of the adiabatic theorem is
the heart of this method.

As is quite natural, the scientific community faced with great skepticism Kieu's method. 
Some tried to find flaws in it and, thus, to show that it cannot actually solve what it claims
it does. Some others, hooked to classical recursion theory, cannot actually accept it that 
what they were doing all these years is actually a special case of a broader yet to be fully 
developed theory and, thus, fiercely  object to the idea that this method can actually solve 
HTP. Warren Douglas Smith~\cite{smith06} has described the ``flaws'' he has managed to 
find in Kieu's method and, thus, believes he has refuted Kieu's plan. However, Kieu does
not really share Smith's objections. Indeed, Kieu~\cite{kieu06a,kieu06b} has shown   
that Smith's objections are actually fallacious. So, do we have a method that can be readilly 
used to solve any Diophantine equation? The answer, in principle, is yes, but it remains to 
see how the individual steps can be effeciently implemented. However, the important thing 
is that this method solves a Turing noncomputable problem, while it does not demand any 
alteration of the physical laws as we have came to know them. 
 
David Malament and Mark Hogarth were probably the first thinkers to seriously propose
the possibility of performing {\em supertasks}, that is, the possibility to perform an 
infinite number of actions in a finite time. In a sense, it is possible to perform a 
supertask: John Earman~\cite[p.~103]{earman95} has noted  that ``an ordinary walk from 
point $A$ to point $B$ involves crossing an infinite number of finite (but rapidly shrinking) 
spatial intervals in a finite time.'' Also, for a similar reason, one may consider that the 
length of any ``natural'' curve (e.g., the coastline of Britain) is not finite 
(see~\cite{peitgen92} for a general discussion of fractal geometry, the mathematical
theory behind such ideas, and~\cite{korvin92} for a more specialized discussion of the 
application of fractal geometry in earth sciences). Naturally, an immediate objection to such
arguments is that it is one thing to estimate the length of a curve and another to perform a
supertask. An intuitive response to this objection is to put forth the idea that since a sequence
of events makes up a curve in spacetime, then it is possible to have a curve with infinite 
length confined within a finite part of spacetime (think of space-filling fractal curves).
Nevertheless, intuition cannot be and should never be a replacement for scientific rigor, 
so I will present real scientific arguments that show why supertasks are indeed plausible.

Roughly, a spacetime is the arena in which all physical events take place. Events
are points in a spacetime that are specified by its time and place. A  world line 
is the unique path of an object as it travels through a spacetime. A Malament--Hogarth 
spacetime is a relativistic spacetime in which a given event $p$ may be preceded by 
a world line $\lambda$ of infinite proper length. In other words, any event $p$ may be 
preceded by an infinite sequence of events. In such a spacetime it is easy to perform 
a supertask: Suppose that we want to see whether Goldbach's conjecture\footnote{Goldbach's 
conjecture states that every even number greater than two is the sum of two primes.} is 
actually true. We can assign to a machine the task of checking all natural numbers against 
Goldbach's conjecture and have an independent observer wait for the machine to finish. 
The machine will operate on the past worldline of the observer and once it has found an 
answer to this problem, it will signal its findings to the observer. Goldbach's conjecture 
is clearly a recursively enumerable problem. More generally, any apparatus 
that can perform a supertask is in principle able to {\em compute} noncomputable problems. 
But, how can we actually implement such an apparatus?

G\'{a}bor Etesi and Istv\'{a}n N\'{e}meti~\cite{etesi02} have proposed  a thought 
experiment that shows how one can perform a supertask within a finite time interval. 
This thought experiment involves the exterior of a Kerr black hole (i.e., a slowly rotating 
black hole), whose inner horizon is actually a Malament--Hogarth spacetime. In particular,
they have proposed a way to construct a computing system that will be able to perform a 
computational supertask. More specifically, they have proposed a system consisting of a 
computer traveling around a Kerr black hole in a stable circular orbit in 
the equatorial plane and a freely falling observer that crosses the outer event horizon 
of the black hole and enters its inner horizon, which is not {\em globally 
hyperbolic},\footnote{A spacetime is globally hyperbolic when it ``does not contain any 
pathologies that prevents the implementation of global Laplacian 
determinism''~\cite[p.~44]{earman95}. In other words,  a globally hyperbolic spacetime is one 
that may accommodate Laplace's demon.} but does not continue into the singularity. 
The computer performs the supertask and the careful observer waits for the result.

Etesi and N\'{e}meti~\cite{etesi02} have discussed  a number of issues related to 
their mental construction, which, however, do not pose any real threat to the 
realization of their system. In particular, initially it was thought that one
should be able to measure time with any imaginable precision. Unfortunately, most
widely accepted models of quantum gravity assume that time and space are not
continuous but rather granular. Nevertheless, Richard Lieu and Lloyd 
W.~Hillman~\cite{lieu03} have given  {\em strong} experimental evidence against the granularity 
of space and time. Thus, this objection is vitiated by the Lieu and Hillman's findings. 
But, Istv\'{a}n N\'{e}meti and Gyula D\'{a}vid~\cite{nemeti06} have concluded that 
it does not really matter whether time is continuous or granular for their set up to work. 
Clearly, it is not that simple to find a black hole, let alone a rotating black hole, 
to check this idea. However, the point is that there is nothing in this mental experiment 
that is physically implausible and so this construction can {\em compute} noncomputable 
problems. 
\section{Information Processing and Life}
Hypercomputation is not physically impossible, but in what 
ways does it affect our understanding of life as a natural phenomenon? To some extend, 
all living beings are information processors and, consequently, their capabilities 
depend partially on the limits of computation. If these limits are not known, then,
clearly, it is not possible to know the limits of the mental capabilities of intelligent 
beings. 

If the ``it from bit'' hypothesis was valid, then (intelligent) living beings would process
information in a discrete manner. However, Michael J.~Spivey and his colleagues have 
reported in~\cite{spivey05} that humans comprehend language in a continuous way. More 
specifically, their conclusion was drawn from observations made during an experiment they 
had carried our with the help of a number of volunteers. This discovery does 
not invalidate the computational metaphor, since one would argue that humans are {\em analog}
computing devices that process language or, more generally, information in a continuous way.
Nevertheless, contra to the commonly accepted view this discovery is an indication, if not
a proof, that information is not discrete in nature. Like light it may have a dual nature,
but is not discrete only. Therefore, one cannot say that information flows in bits, thus
this discovery invalidates the ``it from bit'' hypothesis. In addition, one can safely 
conclude that humans are partially continuous (analog) computing machines that processes
information in a continuous way. However, the question remains: are humans information 
processors only? 

It is true that living organisms interchange information, and this is evident among 
primates. However, to simply state that a flower is blue is usually regarded as 
information with no content, since this has no impact in our life. On the other hand, 
the knowledge that blue flowers may cause some kind of allergy and, at the same time, 
the information that Maria's garden has lots of blue flowers, may prohibit one from visiting 
Maria's house. Generally, humans and (some?) animals perform logical deductions based on 
previously processed information. To say that living creatures are merely information processors, 
it is just like saying that because our brain consists of neurons, brain activity can 
be reduced to neuron firings (see~\cite[p. 185]{baumeister05} for an excellent argument 
against this idea). Also, one has to be really careful with the information he/she 
receives, as, for example, many living creatures use the art of producing unreliable 
information as a way to confuse predators. Thus, information alone is not enough. 
In a nutshell, information is vital for the survival of any living creature, but living 
creatures are not just information processors---they are something more.  But, if living 
creatures are not just information processors, then how does hypercomputation affect 
life as a physical phenomenon?

Hypercomputation and the idea that the human mind has both hypercomputational 
and {\em paracomputational} capabilities, that is, capabilities that go beyond computation, 
are orthogonal. According to this idea our brains are not only more powerful than any 
TM,\footnote{Strictly, speaking even modern digital computers are 
more powerful than TMs as was shown by Peter Wegner (e.g., see~\cite{wegner98}).}
but they can also perform tasks no {\em computing} device can achieve. Clearly, unless one
does not have a concrete description of what is computation, it is not possible to
further argue about the capabilities of the human brain. Computation is symbol-manipulation
process that in addition has some important characteristics.
\begin{definition}\label{comp:def}
Computation is an implementation-independent, systematically interpretable, symbol manipulation
process.~\cite{harnad95}
\end{definition}

It is not difficult to see that certain aspects of any human are not computational
in nature. For example, feelings (e.g., love, wrath), necessities (like the necessity for God),
and even an orgasm are not computational in nature and, therefore, cannot be replicated. 
Even though it is possible to computationally {\em simulate} parts of human behavior.
Interestingly, an anonymous reviewer has pointed out that although an orgasm may have shock 
value to recommend, nevertheless it has not logical foundation. First, one must understand that 
the computational metaphor is about our inability to do anything beyond computation. Therefore, 
if there is something that goes beyond computation, then this will invalidate the computational
metaphor, which is something in agreement with hypercomputation. In addition, this example shows
that we need a broader definition of the terms {\em mechanical} and {\em machine}, since these 
are too closely associated to TMs and their capabilities. These remarks lead to the following 
conclusion. 
\begin{thesis}
Living creatures are biological machines that can process information and have both
hypercomputational and paracomputational capabilities.
\end{thesis}

If our computational capabilities are only delimited by the physical properties of our
world, what is the highest level of intelligence humans will ever reach? First is it
important to say that we cannot give a precise answer to this question, since we do not
know the laws of nature. What we know is an {\em approximation} of the truth, as for
example the laws of Newtonian mechanics are a crude approximation of reality. Another obstacle
is that we do not have at our disposal full models of many important phenomena that
affect our capabilities. Also, it is not known what are the real limits of computation. In
other words, even feasible hypercomputers will not be able to solve some problems, but again
we have no idea what these problems are. Nevertheless, we can dare to predict that humans will 
surpass various milestones one after the other until they will reach some upper limit. This
is not something that can  be achieved in a few years time; it may take thousands, even millions 
of years. But in the end, humans will reach a
\begin{displaymath}
\mbox{homo}\;\underbrace{\mbox{sapiens\ldots sapiens}}_{\mbox{\small $n\gg2$}}
\end{displaymath}
stage in their evolutionary path, where, obviously, $n$ cannot be known, at least for now.
\section{Algorithmic Communication with Extraterrestrial Intelligence?}
In this section I will try to briefly investigate the following question: in the light of
hypercomputation does it make sense to have some sort of algorithmic communication with 
intelligent aliens? Let me start with a remark. It is well-known that for nearly 1,500 years, 
no one could read hieroglyphs, the ancient Egyptian picture-writing. In 1799 a stone with
an inscription in three different languages was unearthed. On this stone the same 
text was written in two Egyptian language scripts (hieroglyphic and Demotic) and in classical Greek. This stone is now known as the Rosetta Stone. This discovery prompted scholars to start
working on the deciphering of the ancient script. Based on earlier work done by other
scholars, Jean-Fran\c{c}ois Champollion had managed to fully understand and decipher the 
hieroglyphic writing. Of course the deciphering of the hieroglyphic writing was possible
because Champollion was fluent in classical Greek. This example from history is very helpful in
understanding first whether it is actually possible to start a communication with 
extraterrestrial intelligence and second what it is required to keep such a communication
alive. 

Roughly speaking, the essence of hypercompution is that computing machines have more
computational power than a TM, but, for the time being, we humans have no idea where
the limits of computation lie. Of course, what any civilization can actually compute with 
their computing devices depends solely on their scientific and technological advancement. 
This simply means that, the more advanced a civilization is, the higher in the arithmetic 
hierarchy lie the entities their computing devices can actually compute. Similar civilizations 
may differ only in the speed their computing devices can deliver computational results. But, 
even this speed up may have really serious effects to our understanding of the cosmos. For
instance, the Mandelbrot set (or "gingerbread man", as it also known in the literature) would 
make no sense to any scientist of, say, the 1950s. These scientists, just like Champollion, 
would need a Rosetta Stone to decipher the ``meaning'' of a computer print-out of the Mandelbrot
set. So this lead us to wonder whether it is reasonable to expect to understand a message sent
by some alien intelligent beings. Unless the alien civilization that tries to communicate with
us has achieved an almost identical technological level, it would be really difficult to communicate
with them. Ideas and ``facts'' known to them may be completely strange to us. For instance,
nowadays scientists really doubt the validity of string theory (see~\cite{woit06} for an
excellent discussion of this matter) and if this theory is actually wrong and some far more
advanced civilization sends us the basic ideas of the theory that actually describes what
string theory is supposed to describe, then clearly no one will be able to understand this 
message. In a nutshell, it is not at all clear whether we are ready to decipher a message from 
an extraterrestrial intelligence that will possible arrive in the near future. Actually, one 
would suggest that we have already received such messages, but maybe we cannot even sense 
their presence! 

On the other side, it does really make sense to try to make our presence known. Of course,
only civilizations that are at same technological level or at a higher technological level,
but not really much higher (modern Copts do not necessarily understand the hieroglyphic writing), 
will be able to possibly understand our messages. Very advanced civilizations may not bother 
getting into the trouble communicating with us. But then again, there is a really small 
possibility that we are actually sending messages to anyone.

\section{Conclusions}
Hypercomputation is a new branch of computer science that is about computing beyond the
CTT. For the time being, we have not constructed any hypercomputer, nevertheless,
it has been shown that hypercomputation is physically plausible.  As a side effect,
hypercomputation affects our understanding of life as a natural phenomenon. Thus,
hypercomputation broadens our understanding of the limits of our intelligence and our
corresponding mental capabilities. In addition, hypercomputation can be used to 
investigate the problem of communication with extraterrestrial intelligence.
The conclusion is that although hypercomputation will affect our understanding of this 
problem, however, this is not enough, since there are other equally important problems 
that have to be taken seriously under consideration. 

In the light of the remarks above, I conclude by saying that hypercomputation as a 
philosophical doctrine and as a scientific discipline has to offer much to our understanding 
of natural phenomena and life in particular. 

{\small
\begin{thebibliography}{10}

\bibitem{baumeister05}
{\sc Baumeister, R.~F.}
\newblock {\em {The Cultural Animal: Human Nature, Meaning, and Social Life}}.
\newblock {Oxford University Press}, Oxford, UK, 2005.

\bibitem{brumfiel07}
{\sc Brumfiel, G.}
\newblock {Quantum leap of faith}.
\newblock {\em {Nature} 446\/} (15 March 2007), 245.

\bibitem{copeland99}
{\sc Copeland, B.~J., and Proudfoot, D.}
\newblock {Alan Turing's forgotten ideas in computer science}.
\newblock {\em {Scientific American} 280\/} (1999), 76--81.

\bibitem{earman95}
{\sc Earman, J.}
\newblock {\em {Bangs, Crunches, Whimpers, and Shrieks}}.
\newblock {Oxford University Press}, New York, 1995.

\bibitem{etesi02}
{\sc Etesi, G., and N\'{e}meti, I.}
\newblock {Non-Turing Computations Via Malament-Hogart Space-Times}.
\newblock {\em {International Journal of Theoretical Physics} 41}, 2 (2002),
  341--370.

\bibitem{gold65}
{\sc Gold, E.~M.}
\newblock {Limiting Recursion}.
\newblock {\em {The Journal of Symbolic Logic} 30}, 1 (1965), 28--48.

\bibitem{harnad95}
{\sc Harnad, S.}
\newblock {Computation Is Just Interpetable Symbol Manipulation; Congnition
  Isn't}.
\newblock {\em {Minds and Machines} 4}, 4 (1995), 379--390.

\bibitem{hintikka97}
{\sc Hintikka, J.}
\newblock {\em {Language, Truth and Logic in Mathematics}}.
\newblock {Kluwer Academic Publishers}, {Dordrecht, The Netherlands}, 1997.

\bibitem{horowitz84}
{\sc Horowitz, E., and Sahni, S.}
\newblock {\em {Fundamentals of Data Structures in Pascal}}.
\newblock {Computer Science Press}, 1984.

\bibitem{kalmar59}
{\sc Kalm{\'a}r, L.}
\newblock {An Argument against the Plausibility of Church's Thesis}.
\newblock In {\em {Constructivity in Mathematics}}, A.~Heyting, Ed.
  {Noth-Holland}, {Amsterdam}, 1959, pp.~72--80.

\bibitem{karthik05}
{\sc Karthik, T.}
\newblock {The Theory of Everything and the future of life}.
\newblock {\em {International Journal of Astrobiology} 3}, 4 (2004), 311--326.

\bibitem{kieu03}
{\sc Kieu, T.~D.}
\newblock {Computing the non-computable}.
\newblock {\em {Contemporary Physics} 44}, 1 (2003), 51--71.

\bibitem{kieu06b}
{\sc Kieu, T.~D.}
\newblock {A Mathematical Proof for a Ground-State Identification Criterion}.
\newblock {Electronic document available from
  \url{http://arxiv.org/quant-ph/0602146}}, 2006.

\bibitem{kieu06a}
{\sc Kieu, T.~D.}
\newblock {On the Identification of the Ground State Based on Occupation
  Probabilities: An Investigation of Smith's Apparent Counterexamples}.
\newblock {Electronic document available from
  \url{http://arxiv.org/quant-ph/0602145}}, 2006.

\bibitem{korvin92}
{\sc Korvin, G.}
\newblock {\em {Fractal Models in the Earth Sciences}}.
\newblock {Elsevier Science Publishers}, Amsterdam, 1992.

\bibitem{kugel86}
{\sc Kugel, P.}
\newblock Thinking may be more than computing.
\newblock {\em Cognition 22\/} (1986), 137--198.

\bibitem{lieu03}
{\sc Lieu, R., and Hillman, L.~W.}
\newblock {The Phase Coherence of Light from Extragalactic Sources: Direct
  Evidence against First-Order Planck-Scale Fluctuations in Time and Space}.
\newblock {\em {The Astrophysical Journal Letters} 585}, 2 (2003), 77--80.
\newblock Electronic version available from
  \url{http://arxiv.org/astro-ph/0301184}.

\bibitem{mendelson63}
{\sc Mendelson, E.}
\newblock {On Some Recent Criticismof Church's Thesis}.
\newblock {\em {Notre Dame Journal of Formal Logic} IV}, 3 (1963), 201--205.

\bibitem{nemeti06}
{\sc N\'{e}meti, I., and D\'{a}vid, G.}
\newblock {Relativistic computers and the Turing barrier}.
\newblock {\em {Applied Mathematics and Computation} 176}, 1 (2006), 118--142.

\bibitem{peitgen92}
{\sc Peitgen, H.-O., J{\"{u}}rgens, H., and Saupe, D.}
\newblock {\em {Chaos and Fractals}}.
\newblock Springer-Verlag, New York, 1992.

\bibitem{perelman02}
{\sc Perelman, G.}
\newblock {The entropy formula for the Ricci flow and its geometric
  applications}.
\newblock Electronic document available from
  \url{http://arxiv.org/abs/math.DG/0211159/}, 2002.

\bibitem{perelman03}
{\sc Perelman, G.}
\newblock {Ricci flow with surgery on three-manifolds}.
\newblock Electronic document available from
  \url{http://arxiv.org/abs/math.DG/0303109/}, 2003.

\bibitem{peter59}
{\sc P{\'e}ter, R.}
\newblock {Rekursivit{\"a}t und Konstruktivit{\"a}t}.
\newblock In {\em {Constructivity in Mathematics}}, A.~Heyting, Ed.
  {Noth-Holland}, {Amsterdam}, 1959, pp.~226--233.

\bibitem{porte58}
{\sc Porte, J.}
\newblock {Quelques Pseudo-Paradoxes de la ``Calculabilit{\`e} Effective''}.
\newblock In {\em {Actes du 2\textsuperscript{me} Congr{\`e}s International de
  Cybernetique, Namur, [Belgium] 26--29 June 1956}}. {Gauthier-Villars},
  {Paris}, 1958, pp.~332--334.

\bibitem{putnam65}
{\sc Putnam, H.}
\newblock {Trial and Error Predicates and the Solution to a Problem of
  Mostowski}.
\newblock {\em {The Journal of Symbolic Logic} 30}, 1 (1965), 49--57.

\bibitem{rogers87}
{\sc Rogers, Jr., H.}
\newblock {\em {Theory of Recursive Functions and Effective Computability}}.
\newblock {The MIT Press}, {Cambridge, MA, USA}, 1987.

\bibitem{sazanov95}
{\sc Sazonov, V.~Y.}
\newblock {On feasible numbers}.
\newblock In {\em {Logic and Computational Complexity, International Workshop
  LCC'94 Indianapolis, IN, USA, October 13–-16, 1994}}, vol.~960 of {\em
  {Lecture Notes in Computer Science}}. {Springer-Verlag}, {Berlin}, 1995,
  pp.~30--51.

\bibitem{smith06}
{\sc Smith, W.~D.}
\newblock {Three counterexamples refuting Kieu's plan for ``quantum adiabatic
  hypercomputation''; and some uncomputable quantum mechanical tasks}.
\newblock {\em {Applied Mathematics and Computation} 178}, 1 (2006), 184--193.

\bibitem{spivey05}
{\sc Spivey, M.~J., Grosjean, M., and Knoblich, G.}
\newblock {Continuous attraction toward phonological competitors}.
\newblock {\em {Proceedings of the National Academy of Sciences of the United
  States of America} 102}, 29 (2005), 10393--10398.

\bibitem{syropoulos08}
{\sc Syropoulos, A.}
\newblock {\em {Hypercomputation: Computing Beyond the Church-Turing Barrier}}.
\newblock Springer, New York, 2008.

\bibitem{wegner98}
{\sc Wegner, P.}
\newblock {Interactive foundations of computing}.
\newblock {\em {Theoretical Computer Science} 192\/} (1998), 315--351.

\bibitem{wheeler90}
{\sc Wheeler, J.~A.}
\newblock {Information, Physics, Quantum: The Search for Links}.
\newblock In {\em {Complexity, Entropy and the Physics of Information}}, W.~H.
  Zurek, Ed. {Westview Press}, {Jackson, Tennessee, USA}, 1990, pp.~3--28.

\bibitem{woit06}
{\sc Woit, P.}
\newblock {\em Not Even Wrong}.
\newblock Jonathan Cape, 2006.

\end{thebibliography}
}
\end{document}